\documentclass[a4paper,11pt]{article}
\usepackage{pos}

\title{A Standard Model explanation for the MiniBooNE anomaly}
\author*[a,b]{Ara Ioannisian}
\author[c]{Carlo Giunti}
\author[d]{Gioacchino Ranucci}

\affiliation[a]{Yerevan Physics Institute,\\
 Alikhanian Brothers \ 2, 375036 Yerevan, Armenia}

\affiliation[b]{Institute for Theoretical Physics and Modeling,\\
375036 Yerevan, Armenia}

\affiliation[c]{Istituto Nazionale di Fisica Nucleare (INFN), Sezione di Torino, \\Via P. Giuria 1, I--10125 Torino, Italy}

\affiliation[d]{Istituto Nazionale di Fisica Nucleare (INFN),
 Sezione di Milano, \\ I--20133 Milano, Italy}

\emailAdd{ara.ioannisyan@cern.ch}
\emailAdd{carlo.giunti@to.infn.it}
\emailAdd{gioacchino.ranucci@mi.infn.it}

\abstract{
We present the results of a new analysis of the data of the MiniBooNE experiment
taking into account the additional background of photons. 
MiniBooNE normalises the rate of photon production to the measured $\pi^0$ production rate. 
We study neutral current (NC) neutrino-induced $\pi^0$/photon production ($\nu_\mu + A \to \nu_\mu +1\pi^0 / \gamma + X$) on carbon nucleus (A=12). Our conclusion is based  on experimental data for photon-nucleus interactions from the A2 collaboration at the Mainz MAMI accelerator. We work in the approximation that decays of the intermediate states (non-resonant N, $\Delta$ resonance, higher resonances) unaffected by its production channel, via photon or Z boson. $1\pi^0+X$ production scales as A$^{2/3}$, the surface area of the nucleus. Meanwhile the photons 
incoherently created in intermediate states decays will leave the nucleus, and that cross section will be proportional to the atomic number of the nucleus. We also took into account the coherent emission of photons. We show that the new photon background can explain part of the MiniBooNE
low-energy excess, thus significantly lowering the number of unexplained MiniBooNE electron-like events from $5.1\sigma$ to $3.6\sigma$.
}

\FullConference{%
  40th International Conference on High Energy physics - ICHEP2020\\
  July 28 - August 6, 2020\\
  Prague, Czech Republic (virtual meeting)
}

\begin{document}
\maketitle

\section{Introduction}
\label{sec:intro}

The MiniBooNE experiment consists of a liquid-scintillator detector
that uses mineral oil (methylene, CH$_2$) to search  for $\stackrel{ {
    (-)}}{\nu_\mu} \to \stackrel{(-)}{\nu_e} $ oscillations. It
reported a significant excess of low-energy (200-500 MeV)
electron-like events in both neutrino
and antineutrino runs  \cite{Aguilar-Arevalo:2018gpe,Aguilar-Arevalo:2020nvw,Aguilar-Arevalo:2013pmq}. This anomaly gave rise to speculations of new physics beyond the Standard Model \cite{Giunti:2019aiy}. 

It is well known that
at these energies MiniBooNE cannot distinguish electrons/positrons
from photons, and must therefore subtract the contribution due to
photoproduction from their measured signal.  They estimate the
photoproduction rate from the NC process $\nu_\mu +CH_2 \to
\nu_\mu+\gamma +X$ using the number of detected $\pi^0$s from $\nu_\mu
+CH_2 \to \nu_\mu+\pi^0 +X$.

We study neutral current (NC) neutrino-induced $\pi^0$/photon production ($\nu_\mu + A \to \nu_\mu +1\pi^0 / \gamma + X$) on carbon nucleus. Our results \cite{Ioannisian:2019kse,Giunti:2019sag} are based  on experimental data for photon-nucleus interactions from the A2 collaboration at the Mainz MAMI accelerator \cite{A2}.
% by the following argument.

\section{The MiniBooNE single photon background}
\label{sec:bck}

In MiniBooNE the main source of single photons is 
$\Delta^{+/0} \to p/n + \gamma$ decay 
of $\Delta^{+/0}$'s produced in neutral-current $\nu_{\mu}$ interactions with the
mineral oil ($\text{C}\text{H}_{2}$) of the detector.
The MiniBooNE estimated this background
through the measurement of $\pi^{0}$'s that are produced by the decay
$\Delta^{+/0} \to p/n + \pi^{0}$,
using the branching fractions~\cite{Tanabashi:2018oca}
\begin{align}
\null & \null
\text{Br}( \Delta^{+/0} \to p/n + \gamma )
=
( 6.0 \pm 0.5 ) \times 10^{-3}
,
\label{Brg}
\\
\null & \null
\text{Br}( \Delta^{+/0} \to p/n + \pi^{0} )
\simeq
2/3
.
\label{Brp}
\end{align}
Final state interactions (FSI) cause the absorption of a fraction of the $\pi^{0}$'s
in the carbon nucleus that was estimated to be about 37.5\%
by the MiniBooNE~\cite{Aguilar-Arevalo:2020nvw}
%\footnote{
%In this paper we do not consider the new MiniBooNE data presented in
%Ref.~\cite{Aguilar-Arevalo:2020nvw}
%because there is still no available data release.}
.
However,
in Ref.~\cite{Ioannisian:2019kse,Giunti:2019sag} it was noted that
measurements of $\pi^{0}$ photoproduction on nuclei~\cite{Krusche:2004uw,Krusche:2004xz}
indicate that the fraction of $\pi^{0}$'s that emerge from the nucleus
and can be observed is given by
\begin{equation}
\dfrac{ N_{\pi^{0}}^{\text{FSI}} }{ N_{\pi^{0}}^{0} }
=
\dfrac{ \sigma_{\text{FSI}}( \gamma + {}^{A}\mathcal{N} \to \pi^{0} + X ) }{ \sigma_{0}( \gamma + {}^{A}\mathcal{N} \to \pi^{0} + X ) }
\simeq
\dfrac{ A^{2/3} }{ A }
=
A^{-1/3}
,
\label{Rpi0}
\end{equation}
where $A$ is the mass number of the target nucleus ${^A}\mathcal{N}$,
$\sigma_{\text{FSI}}$ denotes the measured cross section which includes final state interactions
and
$\sigma_{0}$ denotes the theoretical cross section without final state interactions.
Since the nuclear radius scales approximately as $A^{1/3}$,
the $A^{2/3}$ dependency of
$\sigma_{\text{FSI}}( \gamma + {}^{A}\mathcal{N} \to \pi^{0} + X )$
indicates that only the nuclear surface contributes,
whereas all the $\pi^{0}$ that are produced in the nuclear interior are absorbed~\cite{Krusche:2004zc}.
However,
the uncertainties of the $A^{2/3}$ scaling in Eq.~(\ref{Rpi0}) are not known.

We propose to consider the extreme total absorption
of the $\pi^{0}$ that are produced in the nuclear interior
as indicated by the $\pi^{0}$ photoproduction data~\cite{Krusche:2004uw,Krusche:2004xz}
as the possible cause of an increase of the estimated MiniBooNE single-$\gamma$ background
that can explain at least part of the MiniBooNE low-energy excess.
In other words, we consider Eq.~(\ref{Rpi0}) as an \emph{ansatz}
of the effects of FSI $\pi^{0}$ absorption in a nucleus
that is motivated by the photoproduction data.
The resulting estimate of
the probability of $\pi^0$ escape from the $^{12}\text{C}$ nucleus is
\begin{equation}
\widetilde{S}_{\text{C}}(\pi^0)
\simeq
12^{-1/3}
=
0.437
,
\label{SGIR}
\end{equation}
that is smaller than that estimated
by the MiniBooNE collaboration~\cite{Aguilar-Arevalo:2020nvw},
%\begin{equation}
$S_{\text{C}}^{\text{MB}}(\pi^0)
=
0.625$
.
%\label{SMB}
%\end{equation}

%According to our estimation,
%the number of $\Delta^{+/0}$ produced in neutral-current $\nu_{\mu}$ interactions with $^{12}\text{C}$
%and the number of $\gamma$'s generated by their decay is a factor
%$[\widetilde{S}_{\text{C}}(\pi^0)]^{-1} \simeq 2.3$
%larger than that obtained from the measurement of $\pi^{0}$'s without taking into account FSI.
%This enhancement of the $\Delta \to N \gamma$
%background due to $\pi^{0}$ FSI in the $^{12}\text{C}$ nucleus
%is in approximate agreement with the
%theoretical estimation of a factor about 2.4 in Ref.~\cite{Leitner:2008fg}
%and it is larger than the factor
%$[S_{\text{C}}^{\text{MB}}(\pi^0)]^{-1}=1.6$
%considered by the MiniBooNE collaboration~\cite{Aguilar-Arevalo:2020nvw}.

\begin{table*}[t!]
\renewcommand{\arraystretch}{1.45}
\begin{center}
\begin{tabular}{c|ccc|ccc}
\hline\hline
& \multicolumn{3}{c|}{$\nu$ mode} & \multicolumn{3}{c}{$\bar \nu$ mode}\\
$E^\mathrm{QE}_{\nu}$(GeV) & [0.2,0.3] & [0.3,0.475] & [0.475,1.3] & [0.2,0.3] & [0.3,0.475] & [0.475,1.3]\\
\hline
$f^{\text{th}}_{\text{coh}}$ & 0.09 & 0.13 & 0.06 & 0.16 & 0.16 & 0.07 \\
$f^{\text{th}}_{N^{*}}$ & 0.02 & 0.02 & 0.13 & 0.03 & 0.02 & 0.13 \\
$\widetilde{R}/R_{\text{MB}}$ & 1.52 & 1.56 & 1.61 & 1.62 & 1.61 & 1.62 \\
\hline\hline
\end{tabular}
\end{center} 
\caption{ \label{tab:enhancement}
Estimations of
$f^{\text{th}}_{\text{coh}}$ and $f^{\text{th}}_{N^{*}}$
from Table~2 of Ref.~\cite{Wang:2014nat}
in three ranges of reconstructed neutrino energy $E^\mathrm{QE}_\nu$
in the $\nu$ and $\bar\nu$ modes of the MiniBooNE experiment,
and the corresponding
values of the enhancement factor $\widetilde{R}/R_{\text{MB}}$
that we obtained considering $f^{\text{th}}_{N}\simeq0.1$~\cite{Zhang:2012xn}
and our value (\ref{SGIR}) of
the probability of $\pi^0$ escape from the $^{12}\text{C}$ nucleus.
}
\end{table*}

Moreover,
the MiniBooNE  assumed that
``single gamma events are assumed to come entirely from $\Delta$ radiative
decay''~\cite{Aguilar-Arevalo:2020nvw},
neglecting the additional
contributions to $\gamma$ production from
coherent photon emission,
incoherent production of higher mass resonances,
and incoherent non-resonant nucleon production~\cite{Zhang:2012xn,Wang:2014nat}.
Taking into account these contributions and our value (\ref{SGIR}) of
the probability of $\pi^0$ escape from the $^{12}\text{C}$ nucleus,
the ratio of single-photon events to NC $\pi^{0}$ events is given by
%\begin{widetext}
\begin{equation}
\widetilde{R}
=
\dfrac
{
N_{\text{H}}^{\text{th}}(\Delta \to N \gamma)
+
N_{\text{C}}^{\text{th}}(\Delta \to N \gamma)
+
N_{\text{coh}}^{\text{th}}(\gamma)
+
N^{\text{th}}(N^{*} \to N \gamma)
+
N^{\text{th}}(N \to N \gamma)
}
{\widetilde{N}_{\text{tot}}^{\text{th,obs}}(\pi^0)}
.
\label{Rt1}
\end{equation}
%\end{widetext}
The contributions in the numerator are, respectively,
the theoretically predicted numbers of single-$\gamma$ events due to
$\Delta \to N \gamma$ in $\text{H}$,
$\Delta \to N \gamma$ in $\text{C}$,
coherent photon emission,
incoherent production of higher mass resonances
($N^{*} \to N \gamma$)
and incoherent non-resonant nucleon production
($N \to N \gamma$).
The denominator is the theoretically predicted total number of observed $\pi^0$ events.
Note that only the denominator of Eq.~(\ref{Rt1})
depends on the probability of $\pi^0$ escape from the $^{12}\text{C}$ nucleus,
because a larger escape probability implies a larger number of observed $\pi^0$ events.
The tilde notation indicates that $\widetilde{N}_{\text{tot}}^{\text{th,obs}}(\pi^0)$
corresponds to our value $\widetilde{S}_{\text{C}}(\pi^0)$
in Eq.~(\ref{SGIR}) of such probability.

We can write Eq.~(\ref{Rt1}) as
\begin{align}
\widetilde{R}
=
\null & \null
\dfrac
{
N_{\text{H}}^{\text{th}}(\Delta \to N \gamma)
+
N_{\text{C}}^{\text{th}}(\Delta \to N \gamma)
}
{N_{\text{tot}}^{\text{th,obs}}(\pi^0)}
\left(
\dfrac
{N_{\text{tot}}^{\text{th,obs}}(\pi^0)}
{\widetilde{N}_{\text{tot}}^{\text{th,obs}}(\pi^0)}
\right)
\nonumber
\\
\null & \null
\times
\left( 1 + f^{\text{th}}_{\text{coh}} + f^{\text{th}}_{N^{*}} + f^{\text{th}}_{N} \right)
,
\label{Rt2}
\end{align}
where
$N_{\text{tot}}^{\text{th,obs}}(\pi^0)$
is the total number of observed $\pi^0$ events estimated by the MiniBooNE collaboration
using the probability of $\pi^0$ escape from the $^{12}\text{C}$ nucleus
$S_{\text{C}}^{\text{MB}}(\pi^0)=
0.625$~\cite{Aguilar-Arevalo:2020nvw}.
In Eq.~(\ref{Rt2})
$f^{\text{th}}_{\text{coh}}$,
$f^{\text{th}}_{N^{*}}$, and
$f^{\text{th}}_{N}$
are, respectively,
the theoretically predicted ratios of $\gamma$'s generated
coherently, by higher mass resonances, and non-resonant nucleon production
with respect to those generated by $\Delta$ decay.
In this way, we separated these contributions from the
those generated by $\Delta$ decay that were considered by the MiniBooNE collaboration.

\begin{figure*}[!t]
\centering
\setlength{\tabcolsep}{0pt}
\begin{tabular}{cc}
%\subfigure[]{\label{fig:hst-mbn-bck}
{\includegraphics*[width=0.45\linewidth]{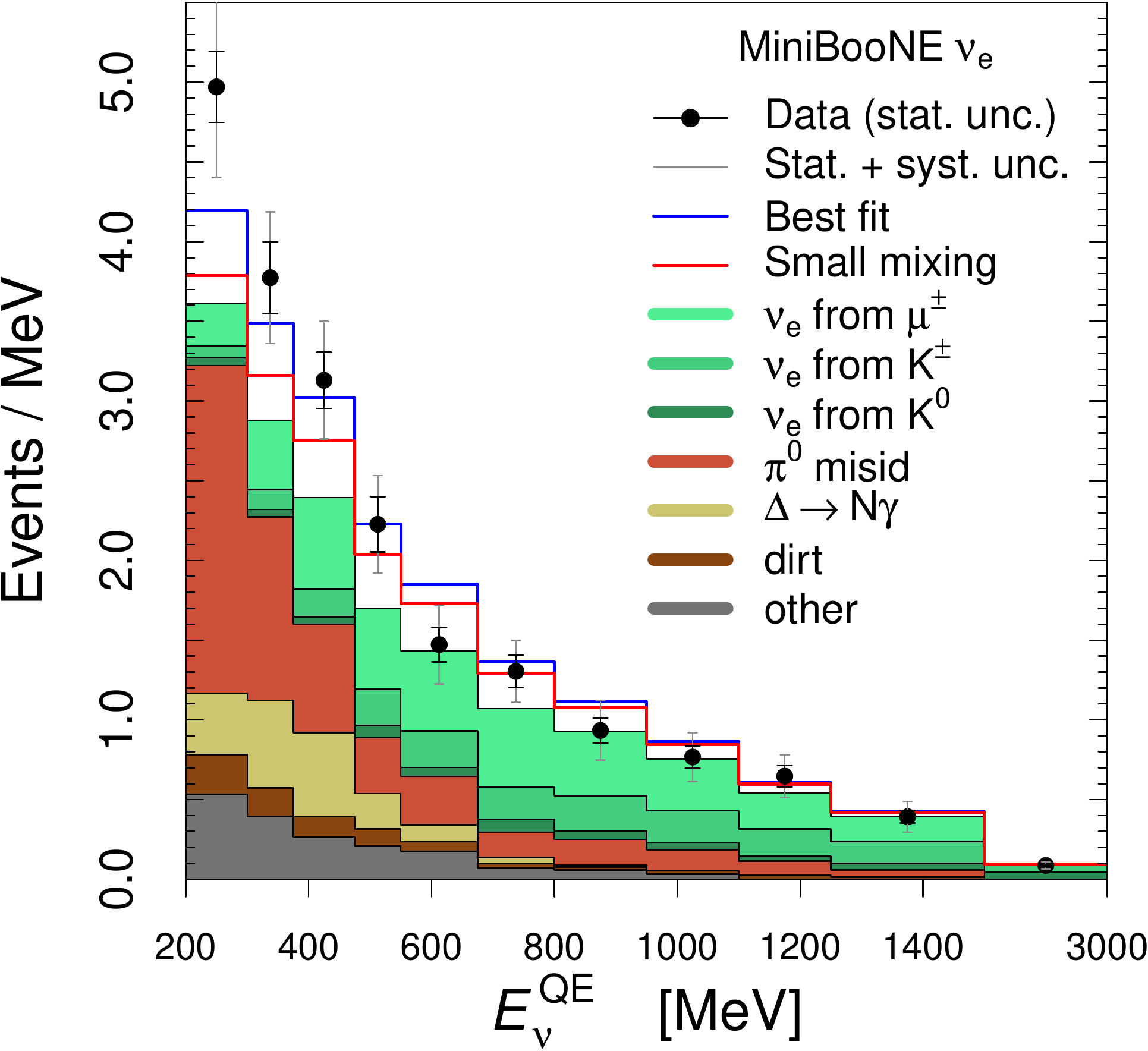}
}
&
%\subfigure[]{\label{fig:hst-mba-bck}
{\includegraphics*[width=0.45\linewidth]{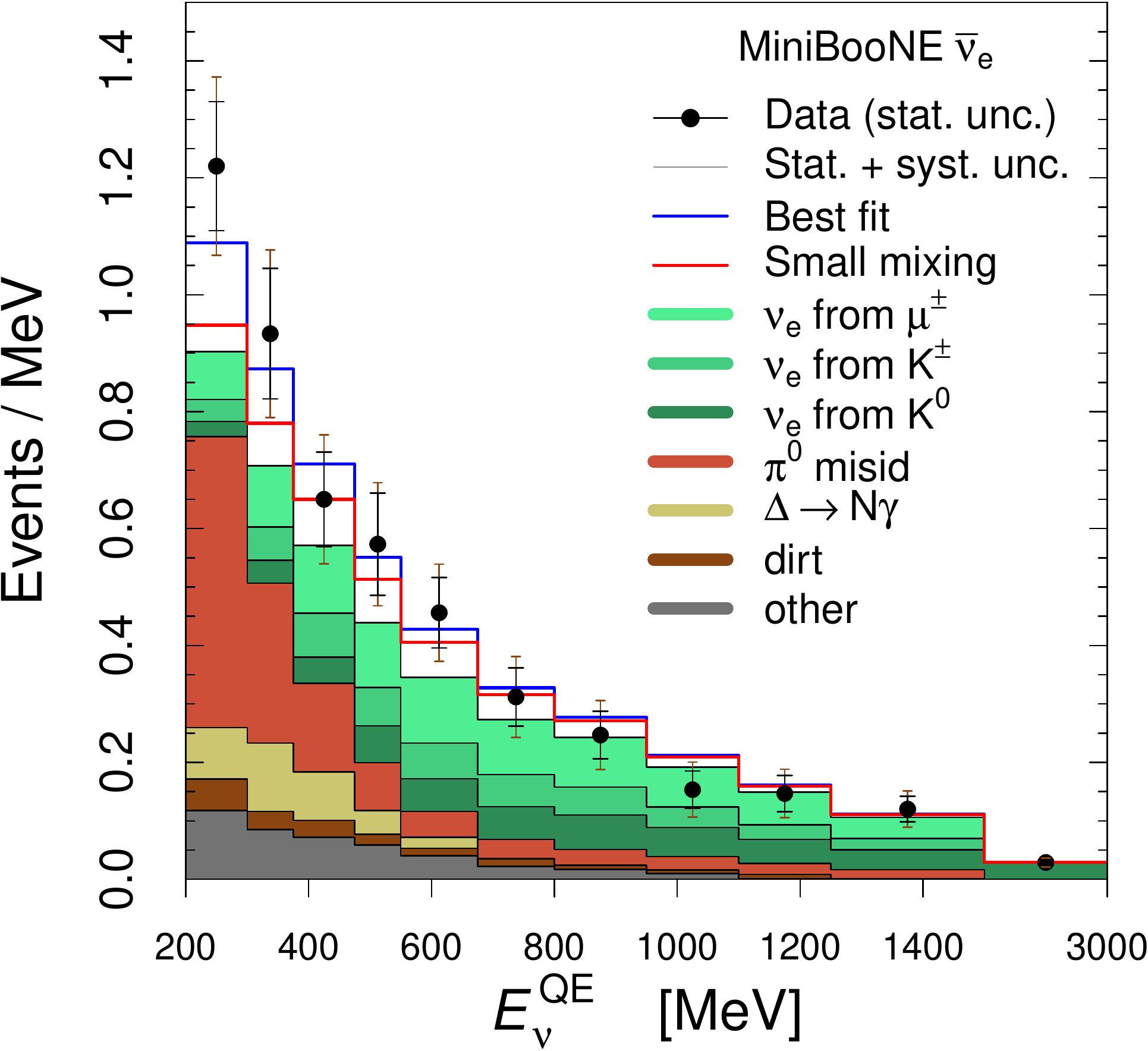}
}
\\
(a) & (b)
\\
&
\\
%\subfigure[]{\label{fig:hst-mbn-bck+fsi}
{\includegraphics*[width=0.45\linewidth]{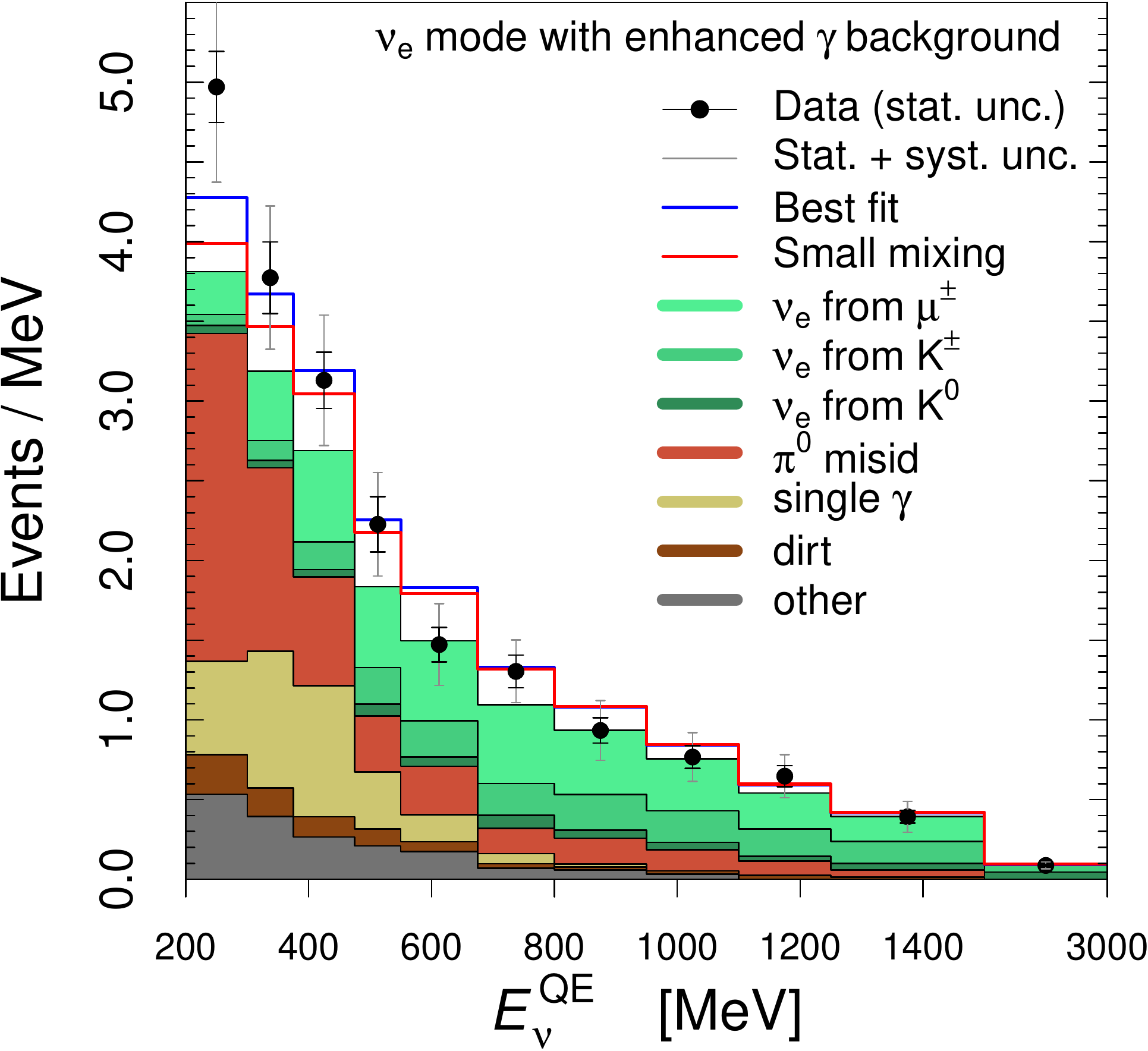}
}
&
%\subfigure[]{\label{fig:hst-mba-bck+fsi}
{\includegraphics*[width=0.45\linewidth]{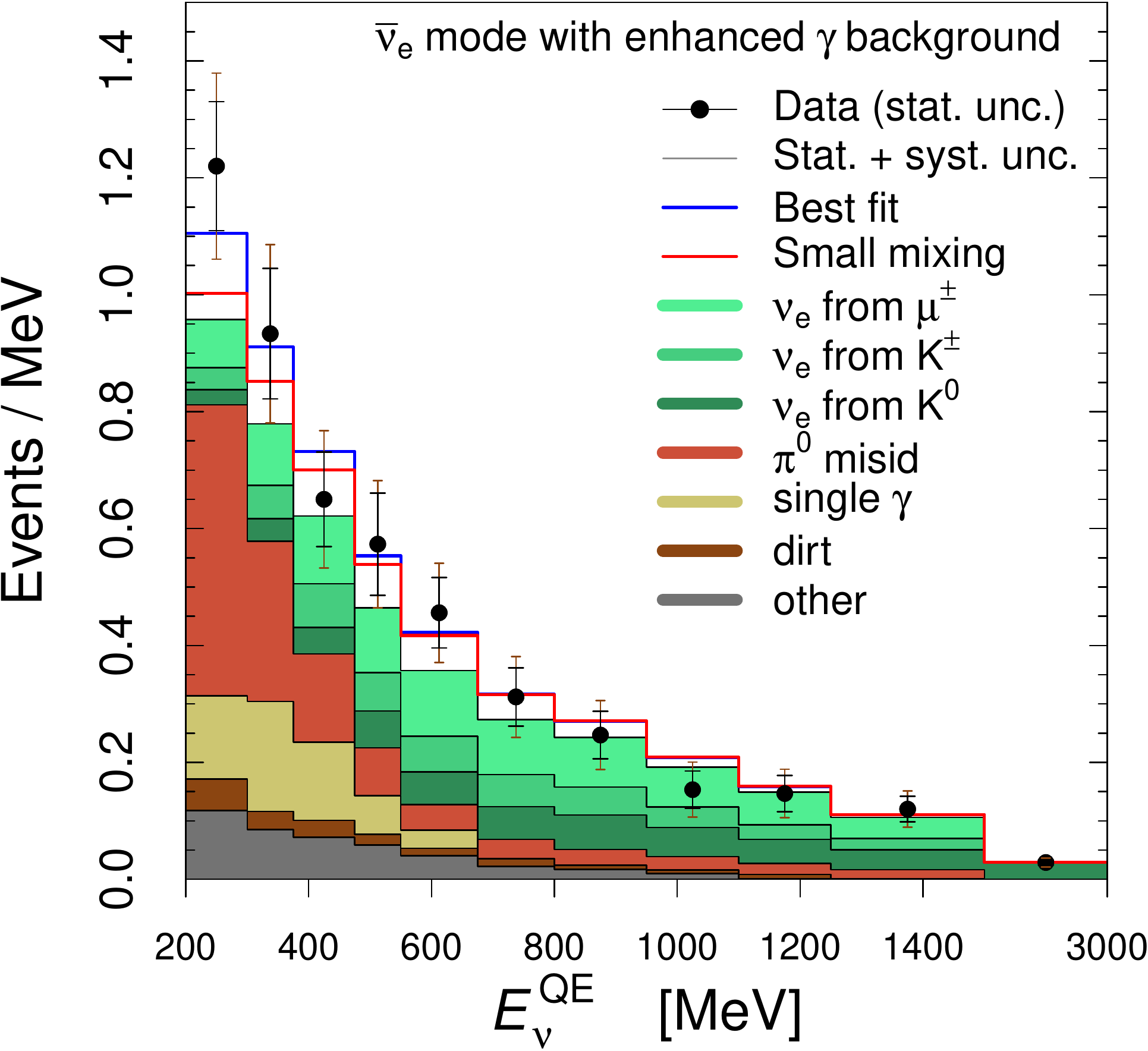}
}
\\
(c) & (d)
\end{tabular}
\caption{ \label{fig:hst}
Comparison of a reproduction of the MiniBooNE electron-like event histograms in (a) 
%\subref{fig:hst-mbn-bck} 
neutrino and (b)
%\subref{fig:hst-mba-bck} 
antineutrino mode 
from Refs.~\cite{Aguilar-Arevalo:2018gpe,Aguilar-Arevalo:2020nvw,Aguilar-Arevalo:2013pmq}
with our versions (c) and (d) 
% \subref{fig:hst-mbn-bck+fsi} and \subref{fig:hst-mba-bck+fsi}
obtained with the enhanced single-$\gamma$ background
due to $A^{1/3}$ $\pi^{0}$ FSI in $^{12}\text{C}$,
coherent photon emission,
incoherent production of higher mass resonances,
and incoherent non-resonant nucleon production.
The blue and red lines show, respectively, the expectations for neutrino oscillations corresponding
to the best fit 
%in Table~\ref{tab:fit} 
(almost maximal mixing) and
the case of small mixing with
$\sin^2\!2\vartheta_{e\mu} = 2.5 \times 10^{-3}$
and
$\Delta{m}^2_{41} = 0.8 \, \text{eV}^2$.
}
\end{figure*}

The first fraction on the right-hand side of Eq.~(\ref{Rt2})
is the ratio of single-$\gamma$ events to NC $\pi^{0}$ events
estimated by the MiniBooNE collaboration:
\begin{equation}
R_{\text{MB}}
=
\dfrac
{
N_{\text{H}}^{\text{th}}(\Delta \to N \gamma)
+
N_{\text{C}}^{\text{th}}(\Delta \to N \gamma)
}
{N_{\text{tot}}^{\text{th,obs}}(\pi^0)}
=
0.0091
.
\label{RMB}
\end{equation}

The second fraction on the right-hand side of Eq.~(\ref{Rt2})
can be calculated by writing it as
\begin{equation}
\dfrac
{N_{\text{tot}}^{\text{th,obs}}(\pi^0)}
{\widetilde{N}_{\text{tot}}^{\text{th,obs}}(\pi^0)}
=
\dfrac
{N_{\text{abs}}^{\text{th,obs}}(\pi^0)+N_{\text{noabs}}^{\text{th,obs}}(\pi^0)}
{\widetilde{N}_{\text{abs}}^{\text{th,obs}}(\pi^0)+N_{\text{noabs}}^{\text{th,obs}}(\pi^0)}
,
\label{Nr1}
\end{equation}
where
$N_{\text{abs}}^{\text{th,obs}}(\pi^0)$
and
$\widetilde{N}_{\text{abs}}^{\text{th,obs}}(\pi^0)$
are the theoretically predicted numbers of observed $\pi^0$ produced in processes
with absorption of $\pi^0$ in the C nucleus,
whereas
$N_{\text{noabs}}^{\text{th,obs}}(\pi^0)$
is the theoretically predicted numbers of observed $\pi^0$ produced in processes
without absorption of $\pi^0$ in the C nucleus.
Note that only
$N_{\text{abs}}^{\text{th,obs}}(\pi^0)$
and
$\widetilde{N}_{\text{abs}}^{\text{th,obs}}(\pi^0)$
depend on the probability of $\pi^0$ escape from the $^{12}\text{C}$ nucleus
and are given by
$
N_{\text{abs}}^{\text{th,obs}}(\pi^0)
=
N_{\text{abs}}^{\text{th,prod}}(\pi^0)
S_{\text{C}}^{\text{th}}(\pi^0)
$
and
$
\widetilde{N}_{\text{abs}}^{\text{th,obs}}(\pi^0)
=
N_{\text{abs}}^{\text{th,prod}}(\pi^0)
\widetilde{S}_{\text{C}}^{\text{th}}(\pi^0)
$.
Therefore,
we can write Eq.~(\ref{Nr1})
as
\begin{equation}
\dfrac
{N_{\text{tot}}^{\text{th,obs}}(\pi^0)}
{\widetilde{N}_{\text{tot}}^{\text{th,obs}}(\pi^0)}
=
\dfrac
{
1
+
\dfrac{N_{\text{noabs}}^{\text{th,obs}}(\pi^0)}{N_{\text{abs}}^{\text{th,obs}}(\pi^0)}
}
{
\dfrac{\widetilde{S}_{\text{C}}^{\text{th}}(\pi^0)}{S_{\text{C}}^{\text{th}}(\pi^0)}
+
\dfrac{N_{\text{noabs}}^{\text{th,obs}}(\pi^0)}{N_{\text{abs}}^{\text{th,obs}}(\pi^0)}
}
.
\label{Nr2}
\end{equation}
We obtained the value of
$
N_{\text{noabs}}^{\text{th,obs}}(\pi^0)
/
N_{\text{abs}}^{\text{th,obs}}(\pi^0)
$
from the contributions to the MiniBooNE $\pi^{0}$
event sample given in Ref.~\cite{Aguilar-Arevalo:2020nvw}:
\begin{equation}
\dfrac
{N_{\text{noabs}}^{\text{th,obs}}(\pi^0)}
{N_{\text{abs}}^{\text{th,obs}}(\pi^0)}
\simeq
0.496
.
\label{Nr3}
\end{equation}
The resulting enhancement factor of the MiniBooNE single-$\gamma$ background is
\begin{equation}
\dfrac{\widetilde{R}}{R_{\text{MB}}}
\simeq
1.25
\left( 1 + f^{\text{th}}_{\text{coh}} + f^{\text{th}}_{N^{*}} + f^{\text{th}}_{N} \right)
.
\label{Rt3}
\end{equation}

The authors of Ref.~\cite{Wang:2014nat} calculated the
number of single photon events from neutral current interactions at MiniBooNE.
From their Table~2 we obtained the estimates of
$f^{\text{th}}_{\text{coh}}$ and $f^{\text{th}}_{N^{*}}$
in Table~\ref{tab:enhancement},
considering three ranges of $E^\mathrm{QE}_\nu$
in the $\nu$ and $\bar\nu$ modes of the MiniBooNE experiment.
For $f^{\text{th}}_{N}$ we considered the 10\% value estimated in Ref.~\cite{Zhang:2012xn}.
Note that we did not use the absolute values of the events calculated in
Refs.~\cite{Zhang:2012xn,Wang:2014nat},
that are in agreement with the MiniBooNE estimates,
and hence in disagreement with our estimations.
We used only the relative values of the $\gamma$'s generated
coherently, by higher mass resonances, and non-resonant nucleon production,
whose estimation can be considered more accurate.

As shown in Table~\ref{tab:enhancement},
we find an enhancement
$\widetilde{R}/R_{\text{MB}}$
of the single-$\gamma$ background in MiniBooNE
by a factor between 1.52 and 1.62 depending on the
energy range and neutrino or antineutrino mode of the MiniBooNE experiment.
This increase of the single-$\gamma$ background
can explain in part the low-energy MiniBooNE excess,
because its largest contribution occur in the lowest energy bins,
as one can see from Figures 1(a) and 1(b) 
that reproduce the MiniBooNE event histograms in
neutrino and antineutrino mode
in Refs.~\cite{Aguilar-Arevalo:2018gpe,Aguilar-Arevalo:2020nvw,Aguilar-Arevalo:2013pmq}.

Figure~\ref{fig:hst} shows a comparison of the standard
MiniBooNE event histograms
(Figures 1(a) and 1(b))
with those obtained with our reevaluation of the single-$\gamma$ background
(Figures 1(c) and 1(d)).
One can see that in the reproductions
1(a) and 1(b)
of the original MiniBooNE histograms
the low-energy bins show a large excess with respect to the background prediction.
The excess is significantly reduced
with our enhanced single-$\gamma$ background.
Only the first energy bin remains with a large visible excess.

\section{Summary}

We have shown that a reassessment of the
single-$\gamma$ background from $\Delta^{+/0}$ decay in the MiniBooNE experiment
taking into account the effect of $A^{1/3}$ $\pi^{0}$ FSI proposed in Ref.~\cite{Ioannisian:2019kse}
and additional contributions to the single-$\gamma$ background
can explain in part the low-energy MiniBooNE excess.
In absence of physics beyond the standard three-neutrino mixing,
our enhanced single-$\gamma$ background
leads to a better fit of the data,
with a goodness-of-fit of $ 2\%$,
with respect to the standard analysis,
that has a goodness-of-fit of $0.02\%$.
The statistical significance of the indication in favour of $\nu_\mu \to \nu_e$ oscillation
decreases from
$5.1\sigma$
to
$3.6\sigma$.

\end{document}